\pdfoutput=1
\documentclass[%
 reprint,
 amsmath,amssymb,
 aps,
pra,
]{revtex4-1}
\usepackage{url}
\usepackage{graphicx}
\usepackage{dcolumn}
\usepackage{bm}
\usepackage{lipsum}
\usepackage{hyperref}
\usepackage{nicefrac}

\newcommand{\expectation}[4][1em]{\langle \makebox{$#2$} \makebox{$|#3|$} \makebox{$#4$} \rangle}
\newcommand{\ket}[1]{| #1 \rangle}
\newcommand{\bra}[1]{\langle #1| }
\newcommand{\MatrixElem}[2]{\langle #1|#2\rangle}
\begin{document}

\preprint{APS/123-QED}

\title{Raman transitions between hyperfine clock states in a magnetic trap}

\author{J B Naber}

\author{L Torralbo-Campo}\altaffiliation[Present address:]{Center for Quantum Science, Physikalisches Institut, Eberhard-Karls-Universit{\"a}t T{\"u}bingen, Auf der Morgenstelle 14, D-72076 T{\"u}bingen, Germany}
\author{T Hubert}%
\author{R J C Spreeuw}
\email{r.j.c.spreeuw@uva.nl}
\affiliation{Van der Waals-Zeeman Institute, University of Amsterdam, Science Park 904, PO Box 94485,
1090 GL Amsterdam, The Netherlands}

\date{\today}

\begin{abstract}
We present our experimental investigation of an optical Raman transition between the magnetic clock states of $^{87}$Rb in an atom chip magnetic trap.
The transfer of atomic population is induced by a pair of diode lasers which couple the two clock states off-resonantly to an intermediate state manifold.
This transition is subject to destructive interference of two excitation paths, which leads to a reduction of the effective two-photon Rabi-frequency.
Furthermore, we find that the transition frequency is highly sensitive to the intensity ratio of the diode lasers.
Our results are well described in terms of light shifts in the multi-level structure of $^{87}$Rb.
The differential light shifts vanish at an optimal intensity ratio, which we observe as a narrowing of the transition linewidth.
We also observe the temporal dynamics of the population transfer and find good agreement with a model based on the system's master equation and a Gaussian laser beam profile.
Finally, we identify several sources of decoherence in our system, and discuss possible improvements.
\end{abstract}

\maketitle


\section{\label{sec:intro}Introduction}
The development of quantum systems with long coherence times is of utmost importance for many applications ranging from atomic clocks \cite{bloom2014optical} to novel quantum information platforms \cite{Cote:2001uo}.
An intrinsic advantage of neutral atoms in this regard is their relatively weak coupling to the environment.
The combination of neutral atoms with Rydberg excitation would yield strong, switchable, interactions over optically resolvable distances, making ultracold atoms increasingly popular with regard to applications in quantum information \cite{Saffman:2010du} and quantum simulation \cite{Weimer:2010ez}.
In this context, several different approaches address the issue of confining atoms:
dipole traps \cite{Grimm:2000kt,Urban:2009vt}, optical lattices \cite{Bloch:2005gn,Schauss:2012ee} and magnetic traps \cite{Gunter:2013fv}.
\par
Our approach is to confine $^{87}$Rb atoms in an array of magnetic microtraps \cite{Leung:2014gw}.
Arrays of magnetically trapped atoms on a chip provide great freedom in the design of trapping geometries, including the integration with other structures on the chip. They also appear attractive in terms of robustness.
The two-dimensional geometry provides intrinsic addressability.
However, magnetic trapping also implies sensitivity to magnetic field fluctuations in the environment, thus opening a source of decoherence.\par

For the alkali metals a possible solution is to use a pair of clock states in the ground state manifold with vanishing linear differential Zeeman shift \cite{Bize:1999gm}.
This pair of states is then a good candidate for a qubit.
For $^{87}$Rb the $5s_{1/2}$ ground state has two different hyperfine-levels $F=1$ and $F=2$ with a $6.8\,$GHz splitting.
Two magnetically trappable states, $\ket{F{=}1,m_F{=}-1}$ and $\ket{F{=}2,m_F{=}1}$, experience the same first order Zeeman-shift at a magnetic field of $B=3.23\,$G, called the ``magic field" \cite{Matthews:1999bb,Harber:2002fy}.
As a result, coherence times of several seconds \cite{Treutlein:2004ft,Deutsch:2010in} have been achieved.
The long coherence time makes $^{87}$Rb atoms on atom chips a candidate for atomic clocks \cite{Szmuk:2015jw}.
\par

This transition can be driven using the combination of a microwave (MW) and a radio-frequency (RF) photon \cite{Lewandowski:2002ji}.
However, this does not allow for spatial addressing of qubits in the sub-millimeter regime.
For addressing it is necessary to use either a fully optical approach or an optically assisted scheme \cite{Weitenberg:2011gn}.
Here we investigate driving this transition with a pair of Raman lasers via an off-resonant excited level
 \cite{Blatt:2008gj,Yavuz:2006gj}.
The Raman excitation lasers are realized by two diode lasers in a heterodyne optical phase-locked loop (OPLL) \cite{Prevedelli:1995tfbaca,Hoeckel:2008ft}.\par

While this constellation can be approximated as a  Lambda type three-level system \cite{Wu:1996bh,Feng:2014gz},
a full description has to involve the multi-level structure of the atom \cite{Bateman:2010hh}.
The multi-level structure must be taken into account to correctly describe light shifts induced by the Raman lasers.
We find that there is an optimum intensity ratio of the two Raman lasers for which the differential light shift of the qubit levels vanishes.
The Raman transition that we study here suffers from a suppression by destructive interference of multiple excitation pathways via different intermediate states \cite{Charron:2006gy}.
As a consequence one cannot reduce the effect of spontaneous emission by going to larger detunings.
Furthermore, we uncover several sources of decoherence in our system, which limit the fidelity of single-qubit operations.
Our observations can be explained by a model accounting for the multi-level structure of the ground and excited state.
Based on this model we discuss possible solutions to increase the fidelity for future experiments based on cold atoms in magnetic traps.

\section{\label{ch:RamanSetup}Experiment}

\begin{figure*}
\includegraphics{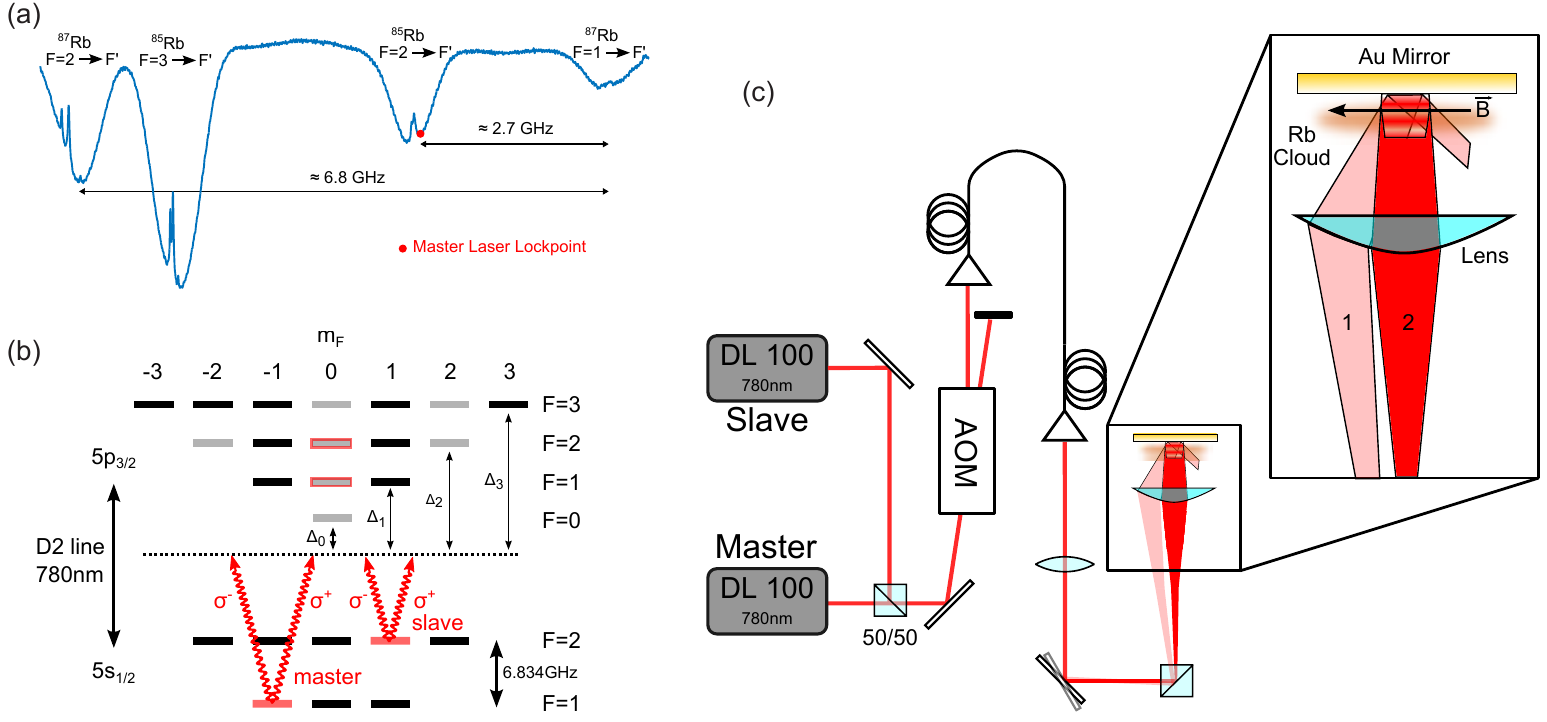}
\caption{\label{fig:Ramansetup}(a) Saturated absorption spectrum in a Rb vapor cell as used for polarization spectroscopy. The hyperfine splitting of the $^{87}$Rb ground state is visible as frequency offset between the two outermost Doppler profiles. The master laser is locked to the $F{=}2\rightarrow F'{=}3$ transition of $^{85}$Rb, effectively introducing a detuning of $\approx2.7\,$GHz for both master and slave laser to the hyperfine manifold of the $5p_{3/2}$ state of $^{87}$Rb. (b) Zeeman manifold of the atomic states involved in the D2 line of $^{87}$Rb. The ground states $\ket{F{=}1,m_F{=}-1}$ and $\ket{F{=}2,m_F{=}1}$ (marked in red) constitute the magnetic clock states at a magnetic field of $B=3.23\,$G. The Raman laser pair has a difference frequency matching the ground state hyperfine splitting of approx. $6.834\,$GHz. With reference to the quantization axis, the laser beams have both $\sigma_+$ and $\sigma_-$ polarization components, introducing two possible transfer paths of atomic populations through $\ket{F'{=}1,m_{F'}{=}0}$ and $\ket{F'{=}2,m_{F'}{=}0}$ (red border). In addition, coupling to several other states (marked in gray) induces AC stark shifts . (c) Sketch of the optical setup. The master and slave laser beams are overlapped, guided through an AOM for pulse generation and coupled into an optical fiber. After the fiber, the laser beam is expanded by a lens and guided to the atom chip experiment. A lens focuses the laser beam to a $1/e^2$ radius of $100\,\mu$m at the atomic cloud, which is confined at $30\,\mu$K in a magnetic z-wire trap with the Ioffe-field pointing parallel to the chip surface. A change in the beam path leads to the two configurations labeled 1 and 2, where the first corresponds to a travelling wave and the latter to a standing wave induced by the reflection of the atom chip.}
\end{figure*}

\subsection{\label{ch:trap}The magnetic trap}
Our experiments are performed by optical excitation of $^{87}$Rb atoms confined in a magnetic Ioffe-Pritchard (IP) type trap formed by a z-shaped wire. 
This wire is located behind a multi-layer atom chip, featuring a layer of micro-structured FePt for the creation of magnetic microtraps \cite{Leung:2014gw}. 
These microtrap potentials are not relevant for the results shown in this paper as the experiments are performed here at distances of more than $100\,\mu$m to the chip surface. 
Initially, the atoms are loaded from a background vapor of $^{87}$Rb into a mirror magneto-optical trap (MOT), where an Au coating on the chip surface acts as the mirror.
The atoms are optically pumped  into the $|F,m_F\rangle=|1,-1\rangle$ magnetic Zeeman level of the $5s_{1/2}$ ground state \cite{HAPPER:1972uq}, which is subsequently trapped in the IP magnetic trap. 
The atomic cloud is cooled in the IP trap by forced RF evaporation to a temperature of $\sim 30\,\mu$K.\par
After the evaporation $10^5$ atoms are left in the magnetic trap.They are detected by absorption imaging using a laser resonant to the $5s_{1/2},F{=}2\rightarrow 5p_{3/2},F'{=}3$ transition. 
With respect to the quantization axis defined by the Ioffe field [see Fig.~\ref{fig:Ramansetup}(c)] the laser light is $\pi$-polarized.
Atoms transferred to the $|F, m_F\rangle=|2,1\rangle$ state via the Raman transition are thus detected, while the population remaining in $|F, m_F\rangle=|1,-1\rangle$ is not visible in the absorption images.
As an independent measure of overall atom number, we can transfer atomic population from the $F=1$ to the $F=2$ ground state by using an independent laser on the transition $F{=}1\rightarrow F'{=}2$ before the imaging.
\subsection{\label{ch:optics}Optical setup}
At the heart of the optical setup there is a heterodyne optical phase-locked loop (OPLL) stabilizing the relative phase and frequency of two diode lasers at $780\,$nm.
This setup is described in greater detail in reference \cite{Naber:2015wk}.
The two Raman lasers, called ``master laser'' and ``slave laser'' according to their role in the locking-scheme, are offset in frequency by $\sim 6.834\,$GHz.
This frequency offset is generated by down-mixing the beat signal of the two lasers with the signal of a commercial microwave generator, and matches the hyperfine splitting of the ground state. 
The master laser is locked to the $F{=}2\rightarrow F'{=}3$ transition of $^{85}$Rb by means of polarization spectroscopy \cite{Pearman:2002kb} in a Rb vapor cell, effectively introducing a fixed detuning of  about $-2.7\,$GHz for both master and slave laser to the hyperfine manifold of the $5p_{3/2}$ state of $^{87}$Rb [see Fig.~\ref{fig:Ramansetup}(a)].
Here the master laser operates at a higher frequency, aiming at the atomic population in $|1,-1\rangle$.
Throughout the paper we define red detunings as $\Delta_i<0$.
The relative frequency of slave and master laser, denoted $\delta$, can be computer-controlled to a Hz precision by changing the frequency of the microwave generator.\par

The light of both lasers is overlapped and then guided through an acousto-optical modulator (AOM), the first diffraction order of which is coupled into a polarization maintaining optical fiber [see Fig.~\ref{fig:Ramansetup}(c)].
The AOM is used for generating square shaped pulses down to $50\,$ns length. 
After the fiber, the laser light is guided to the atom chip experiment, passing through a lens of $f=75\,$mm focal length which renders the beam slightly divergent.
The beam then reaches the in-vacuum imaging lens (NA=0.4, $f=19\,$mm, for a detailed description see reference \cite{Leung:2014gw}), and is narrowed down to a waist ($1/e^2$ beam radius) of $\sim 100\,\mu$m at the atomic cloud (the minimal waist lies behind the chip). 
The beam alignment is altered to yield two different paths labeled 1 and 2 as shown in Fig.~\ref{fig:Ramansetup}(c).  
Beam path 1 is arranged such that the retro-reflected beam does not hit the atomic cloud.
In contrast, the beam in path 2 is reflected back into the cloud and induces a standing wave pattern at the overlap with the incoming beam.
As both beams propagate almost perpendicular to the Ioffe-axis of the trap, they contain $\sigma^+$ and $\sigma^-$ polarization components relative to the quantization axis as determined by the magnetic field.
\par
 
Spectroscopic measurements are performed by scanning the relative frequency of the Raman laser pairs and taking an absorption image of the cloud. In every experimental cycle (duration $\sim 20\,s$) the atoms are loaded in the magnetic trap in the $|F{=}1,m_F{=}{-}1\rangle$ state. 
They are then exposed to a well-defined pulse of the Raman laser light, inducing population transfer to the $|F{=}2,m_F{=}1\rangle$ state. Atoms in this state are detected by the imaging laser.

\section{\label{ch:model}Theory}
\subsection{\label{ch:three}Three level approximation}
The master and slave laser are far-detuned (about $-2.5\,$GHz) from the hyperfine manifold of the $5p_{3/2}$ level.
If their relative frequency matches $\Omega_{01}$, the energy splitting between $|0\rangle=|F{=}1, m_F{=}{-}1\rangle$ and $|1\rangle=|F{=}2, m_F{=}1\rangle$, atomic population can undergo transfer between the two states. 
Excited states which are off-resonantly coupled by the Raman lasers serve as a virtual level for this transition.
As the total change in angular momentum is $\Delta m_F=+2$, the transition must involve the $\sigma^+$ component of the master laser and the $\sigma^-$ component of the slave laser.
Two states which can act as an intermediate state for the Raman transition are $|e1\rangle=|F'{=}1, m_{F'}{=}0\rangle$ and $|e2\rangle=|F'{=}2, m_{F'}{=}0\rangle$ from the excited state manifold [see Fig.~\ref{fig:Ramansetup}(b)].\par

In the state space defined by $|0\rangle$, $|1\rangle$ and $|e1\rangle$ (and similarly for $|e2\rangle$), the levels are coupled with Rabi frequencies $\Omega_{0,e1}=\nicefrac{1}{\hbar}\,\expectation{e1}{\bm{\hat{d}}\cdot\bm{E}_M(\bm{r}_\mathrm{at})}{0}$ and $\Omega_{1,e1}=\nicefrac{1}{\hbar}\,\expectation{e1}{\bm{\hat{d}}\cdot\bm{E}_S(\bm{r}_\mathrm{at})}{1}$.
Here $\bm{\hat{d}}$ denotes the atomic dipole operator, and $\bm{E}_{M/S}(\bm{r}_\mathrm{at})$ is the electric field of the master and slave laser respectively at the atom location $\bm{r}_\mathrm{at}$.
Using the rotating wave approximation (RWA) and adiabatically eliminating $|e1\rangle$ (which is a valid assumption if $|\Delta_1|\gg \Omega_{0,e1},\Omega_{1,e1}$), the system reduces to an effective two-level system, where $|0\rangle$ and $|1\rangle$ are coupled by the two-photon Rabi frequency $\Omega_1=\Omega_{0,e1}\Omega_{1,e1}/(2\Delta_1)$. 
Likewise, the same analysis for $|0\rangle$, $|1\rangle$ and $|e2\rangle$ yields the Rabi frequency $\Omega_2=\Omega_{0,e2}\Omega_{1,e2}/(2\Delta_2)$.\par
We have to add the contributions from both excitation paths established by $|e1\rangle$ and $|e2\rangle$ coherently.
To calculate the matrix elements, we use the dipole operator in the circular basis,
$\hat{\mathbf{d}}=e\,(\hat{r}_{-1},\hat{r}_{0},\hat{r}_{1})$, with $r_{0}=z$ and $r_{\pm1}=(x\pm iy)/\sqrt{2}$.
By applying the Wigner-Eckart theorem we calculate the matrix element between ground $|F m_F\rangle$ and excited $|F^\prime m_{F^\prime}\rangle$ states as 
\begin{equation}
\label{eq:wigner}
\expectation{F^\prime m_F^\prime}{\hat{\mathbf{d}}}{Fm_F}=\expectation{F^\prime}{|\hat{\mathbf{d}}|}{F}\langle F m_F\,1\, q|F^\prime m_{F^\prime}\rangle.
\end{equation}
The term $\expectation{F^\prime}{|\hat{\mathbf{d}}|}{F}$ represents the reduced matrix element, and $\langle F m_F\,1\, q|F^\prime m_{F^\prime}\rangle$ is the Glebsch-Gordan coefficient describing the coupling of different magnetic sublevels for $q\in\{-1,1\}$.
For the Rubidium D2 line with transition frequency $\omega_0=2\pi c/\lambda$ and natural linewidth $\Gamma/2\pi=6\,$MHz, the reduced matrix element can be written as $\expectation{F^\prime}{|\hat{\mathbf{d}}|}{F}=Dd_{FF'}$, with $D=\sqrt{3\pi\varepsilon_0 c^3 \hbar \Gamma/\omega_0^3}$.
The unitless coefficient $d_{FF'}$ is a function of the angular momentum quantum numbers $F$ and $J$ of the states involved.\par

As $\expectation{1}{r_{-1}}{e1} \expectation{0}{r_{1}}{e1}=-\expectation{1}{r_{-1}}{e2} \expectation{0}{r_{1}}{e2}$, we see that the two excitation pathways interfere destructively.
The Rabi frequency of the Raman transition is given by \cite{Wu:1996bh}
\begin{equation}
\label{eq:rabi}
\Omega_R=\frac{\Omega_{0,e1}\Omega_{1,e1}}{2\Delta_1}+\frac{\Omega_{0,e2}\Omega_{1,e2}}{2\Delta_2}=\frac{\Omega_{0,e1}\Omega_{1,e1}}{2\Delta_1}\mathfrak{D}(\Delta_1),
\end{equation}
with $\mathfrak{D}(\Delta_1)=(\Delta_2-\Delta_1)/\Delta_2=\Delta_{21}/(\Delta_1+\Delta_{21})$. The term $\mathfrak{D}(\Delta_1)$ is effectively a reduction resulting from the destructive interference of the two excitation paths. For our choice of detuning, $\Delta_1=-2\pi\times 2290\,$MHz, and $\Delta_{21}=-2\pi\times 157\,$MHz, we get $\mathfrak{D}(\Delta_1)\approx 0.07$. 
Furthermore, for large detuning $|\Delta_1 | \gg |\Delta_{21}|$ the destructive interference essentially changes the scaling of $\Omega_R$ to $\sim 1/\Delta_{1}^{2}$. 
\subsection{\label{ch:shifts_th}Light shifts}
Besides being coupled to the states $|e1\rangle$ and $|e2\rangle$, the ground states $|0\rangle$ and $|1\rangle$ are also off-resonantly coupled to additional excited states. Due to the master and slave laser polarization, in total 7 states (denoted $\mathfrak{S}$) are coupled to the ground states [states labeled gray in Fig.~\ref{fig:Ramansetup}(b)]. 
These couplings induce light shifts of the ground state atomic energies.
The master laser couples $|0\rangle$ and the slave laser $|1\rangle$ to the excited states, with $\Delta_0\ldots\Delta_3$ as the relevant detunings. 
It should be noted that there are non-negligible contributions from light shifts due to the master laser on $|1\rangle$, and from the slave laser on $|0\rangle$ as well. The relevant detunings are just given by the sum of $\Delta_0\ldots\Delta_3$ and the ground state hyperfine-splitting.\par

As the lasers are far off-resonant, the light shifts can be described as a second-order perturbation to the ground state energies \cite{Grimm:2000kt}. For either of the ground states $|g\rangle \in G=\{|0\rangle,|1\rangle\}$, the shift in energy can be approximated as
\begin{eqnarray}
\label{eq:shifts}
\frac{\Delta E_g}{\hbar}=\frac{\Gamma^2}{8}\frac{I_M}{I_\text{sat}}\sum\limits_{\substack{l\in\mathfrak{S}\\q={-}1,1}}\frac{|d_{F_gF_l^\prime}\varepsilon_q^M\langle F_gm_{F_g}\,1\, q|
F^\prime_l m_{F_l^\prime}\rangle|^2}{\Delta_{lg}^M}\nonumber\\
+\frac{\Gamma^2}{8}\frac{I_S}{I_\text{sat}}\sum\limits_{\substack{l\in\mathfrak{S}\\q={-}1,1}}\frac{|d_{F_gF_l^\prime}\varepsilon_q^S\langle F_gm_{F_g}\,1\, q|
F^\prime_l m_{F_l^\prime}\rangle|^2}{\Delta_{lg}^S}.
\end{eqnarray}
Here $I_M,I_S$ is the intensity of master and slave laser, $I_\text{sat}=1.6\,\mathrm{mW}/\mathrm{cm}^2$ is the saturation intensity of the $F{=}2\rightarrow F^\prime{=}3$ transition, and $\Delta_{lg}^M,\Delta_{lg}^S$ are the respective detunings of the lasers.
With $\varepsilon_q^{M(S)}$ we denote the polarization components of the master (slave) laser and set $\varepsilon_1^{M(S)}=\varepsilon_{-1}^{M(S)}=\sqrt{\nicefrac{1}{2}}$, $\varepsilon_0^{M(S)}=0$.
\begin{figure}
\includegraphics{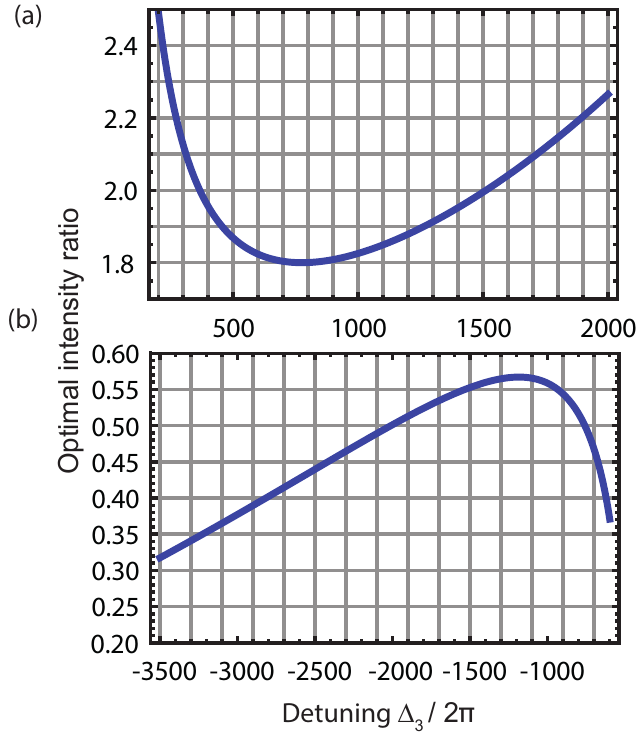}
\caption{\label{fig:ratios} Optimal intensity ratio $I_M/I_S$ of master and slave laser as a function of detuning $\Delta_3$ from $5p_{3/2},F=3$ for positive (a) and negative (b) detuning. The optimal ratio is defined by the intensity ratio at which the differential light shift $\Delta E =\Delta E_1-\Delta E_0$ given by Eq. \eqref{eq:shifts} vanishes.}
\end{figure}
In Eq. \ref{eq:shifts} we neglect the off-diagonal components in the light-shift hamiltonian. This is a good approximation as long as the two-photon Rabi frequency is small compared to the difference in Zeeman shifts of the corresponding $m_F$ states. 
Evaluating the differential light shift between the ground states, $\Delta E =\Delta E_1 - \Delta E_0$, we find that $\Delta E$ is a bilinear function $a I_M - b I_S$ of the two laser intensities. The coefficients $a,b$ are rational functions of the detuning $\Delta_3$ and the overall intensity as follows from Eq. \eqref{eq:shifts}. 
One can thus find an optimal ratio $R=I_M/I_S$ for a given detuning to the excited state, for which the differential light shift vanishes. This optimal ratio is plotted for both positive and negative detunings expressed as $\Delta_3$ in Fig.~\ref{fig:ratios}.
In our experiment, $\Delta_3=-2\pi\times2792\,$MHz, for which an optimal ratio of $R=0.40$ is predicted by Eq. \eqref{eq:shifts}.

\subsection{\label{ch:full}Off-resonant scattering and trap loss}
The off-resonant coupling introduced by the Raman lasers also induces off-resonant scattering from the excited state manifold $\mathfrak{S}$ [states labeled gray in Fig.~\ref{fig:Ramansetup}(b)]. 
This results in atomic population transfer within the ground state manifold, leading to decoherence on the Raman transition and loss of atoms to other magnetic sublevels.
In the low saturation limit ($|\Delta| \gg \Omega$), we can write the scattering rate from $|g_i\rangle=|F_i m_{F_i}\rangle \in G$ to another ground state $|g_f\rangle=|F_f m_{F_f}\rangle$ as the sum of all possible scattering paths through excited states $|l\rangle \in \mathfrak{S}$ induced by the Raman lasers:

\begin{multline}
\label{eq:scattering}
R_{g_i\rightarrow g_f}=\frac{\Gamma^3}{8 I_\text{sat}}\times\\
\Bigg(I_M\bigg|\!\!\!\!\!\sum\limits_{\substack{l \in \mathfrak{S}\\ q_1={-}1,1}}\!\!\!\!\!\frac{d_l\varepsilon_{q_1}^M\MatrixElem{F_f m_{F_f} 1 q_2}{F_l^\prime m_{F_l^\prime}}\MatrixElem{F_i m_{F_i} 1 q_1}{F_l^\prime m_{F_l^\prime}}}{\Delta_{lg_i}^M}\bigg|^2\\
+I_S\bigg|\!\!\!\!\!\sum\limits_{\substack{l \in \mathfrak{S}\\ q_1={-}1,1}}\!\!\!\!\!\frac{d_l\varepsilon_{q_1}^S\MatrixElem{F_f m_{F_f} 1 q_2}{F_l^\prime m_{F_l^\prime}}\MatrixElem{F_i m_{F_i} 1 q_1}{F_l^\prime m_{F_l^\prime}}}{\Delta_{lg_i}^S}\bigg|^2\Bigg).
\end{multline}
In this expression, the parameter $q_2$ obeys $q_2=m_{F_l}-m_{F_f}$ and we write $d_l=d_{F_iF_l^\prime}d_{F_fF_l^\prime}$.
Note that for each laser we use the coherent sum of the off-resonant scattering amplitudes \cite{Cline:1994kw}.\par

The off-resonant scattering populates all ground states, of which only $|F{=}1,m_F{=}{-1}\rangle$, $|F{=}2,m_F{=}1\rangle$ and $|F{=}2,m_F{=}2\rangle$ are magnetically trappable.
Thus atoms can leave the trap by optical pumping into non-trappable states.
The escape rate is related to the magnetic trap frequencies.
In order to include these effects in the description of the Raman transition, we describe the dynamics in terms of the density matrix $\rho$ of the ground state manifold $\mathcal{G}$.
Then the time evolution of the system is governed by the master equation
\begin{equation}
\label{eq:master}
\frac{d}{dt}\rho=-\frac{i}{\hbar}[H_\text{coh},\rho]+\mathcal{L}_\text{scat}(\rho)+\mathcal{L}_\text{loss}(\rho).
\end{equation}
The operator $H_\text{coh}$ describes the coherent dynamics of the system, incorporating the Raman coupling between states $\ket{0}$ and $\ket{1}$ given by Eq. \eqref{eq:rabi}, the light shifts given by Eq. \eqref{eq:shifts} and the Zeeman energy $E=\mu_Bg_Fm_FB$. The magnetic field $B$ is chosen to be the magic field $B=3.23\,$G, where the differential linear Zeeman shift between $\ket{0}$ and $\ket{1}$ vanishes.
The Lindblad super-operator $\mathcal{L}_\text{scat}(\rho)$ describes the decoherence induced by the off-resonant scattering:
\begin{equation}
\label{eq:lindblad}
\mathcal{L}_\text{scat}(\rho)=\frac{1}{2}\sum\limits_j(2c_j\rho c^\dagger_j-c_j^\dagger c_j \rho - \rho c_j^\dagger c_j).
\end{equation}
The sum runs over all possible combinations of $j=(\ket{g_i},\ket{g_f})$ which are coupled by Eq.~\eqref{eq:scattering}. The jump operators $c_j$ are given by $c_j=\sqrt{R_{g_i\rightarrow g_f}}\;\ket{g_f}\bra{g_i}$.
The third term in Eq.~\eqref{eq:master} describes the trap loss induced by the atomic population in magnetically non-trappable states.
We can express this as 
\begin{equation}
\label{eq:loss}
\mathcal{L}_\text{loss}(\rho)=-\frac{\gamma_\text{loss}}{2}\sum\limits_j(P_j \rho + \rho P_j ),
\end{equation}
where the sum includes all non-trappable states $\ket{j}$ and $P_j$ is the projector on state $\ket{j}$.
In this expression, $\gamma_\text{loss}$ is the effective loss rate from the trap, here assumed to be independent of $j$.\par

As a low decoherence rate is desired, it should be noted that lower scattering rates can usually be obtained using a larger detuning $\Delta$. According to Eq. \eqref{eq:scattering}, the scattering rate scales as $R \propto 1/\Delta^2$. However, due to the two-path interference the Rabi-frequency $\Omega_R$ given in Eq. \eqref{eq:rabi} scales as $\Omega_R \propto 1/\Delta^2$ as well for large detunings ($|\Delta| \gg \Delta_{12}$). Therefore, increasing the detuning does not solve the problem of scattering induced decoherence.
\section{\label{ch:results}Experimental Results}
For each absorption image taken after the excitation, which detects the atoms in the $|F{=}2,m_F{=}1\rangle$, we fit a two-dimensional Gaussian function to the atomic cloud. From that, we extract the peak optical density which is plotted against the relative frequency of the Raman lasers.\par
The Raman transition includes the absorption and stimulated emission of a photon via the virtual level. 
As the frequencies of the two Raman lasers are almost identical, we expect that we obtain a Doppler-free spectrum for two co-propagating (co-p.) beams.
This is opposite to the situation in saturated absorption spectroscopy, where counter-propagating (cn-p.) pump and probe beams yield Doppler-free features.
As a proof of principle, we observe the spectrum for beam alignment 1 [compare Fig.~\ref{fig:Ramansetup}(c)], which corresponds to the co-p. case.
We obtain a spectrum which is well described by a Lorentzian function with FHWM of $8.8\pm 1.0\,$kHz.\par
We then switch to alignment 2, which contains also a counter-propagating component because of the reflected beam.
\begin{figure}
\includegraphics{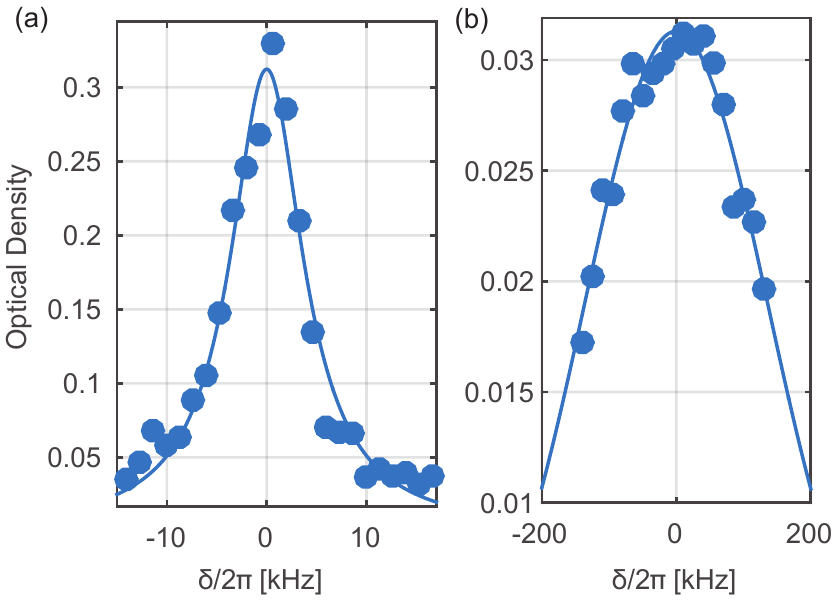}
\caption{\label{fig:doppler} (a) Raman spectrum obtained for the beam path 1 (co-propagating) as depicted in Fig.~\ref{fig:Ramansetup}. The curve shows the fit of a Lorentzian function with a FWHM of $8.8\,$kHz. (b) Spectrum obtained for beam path 2 (counter-propagating), where the incoming and reflected laser beams partially overlap. The spectrum is clearly broadened as compared to (a), indicating Doppler-broadening. The curve is based on a Gaussian function with FWHM of $321\,$kHz. Note also the change in vertical scale for the two figures.}
\end{figure}
The spectrum is clearly broadened as compared to alignment 1.
Additionally, the amount of transferred atoms is largely reduced.
For the Gaussian function fitted to the spectrum in Fig.~\ref{fig:doppler}, we find a FWHM $\Delta f=321\pm 24\,$kHz. 
Following reference \cite{Bjorkholm:1976ba}, the linewidth of a far off-resonant two-photon Raman transition ($\Delta f \ll \Delta$) for two counter-propagating beams is related to the temperature $T$ of an atomic gas by $\Delta f = (2f_R)\sqrt{8k_BT \ln 2/(m_\text{Rb}c^2)}$. 
Neglecting the difference between the Raman lasers' frequency, $f_R$ represents either the slave or master laser frequency. 
For our value of $\Delta f$, we find a temperature of $T=29\pm 4\,\mu$K, in good agreement with an independent measurement of the cloud temperature by RF spectroscopy.\par  

In all subsequent measurements we choose the Doppler-free alignment 1.
\subsection{\label{ch:ratios}Intensity ratio and light shifts}
From our evaluation in section \ref{ch:shifts_th} we expect to see an influence of the intensity ratio of master and slave laser on the spectrum.
Notably, the relative frequency between $\ket{0}$ and $\ket{1}$ should be a linear function of that ratio. Furthermore, at a specific ratio the relative frequency shift is supposed to vanish.
This ratio is $I_M/I_S=0.40$ for the detuning $\Delta_3$ chosen in our experiment.\par
\begin{figure}
\includegraphics{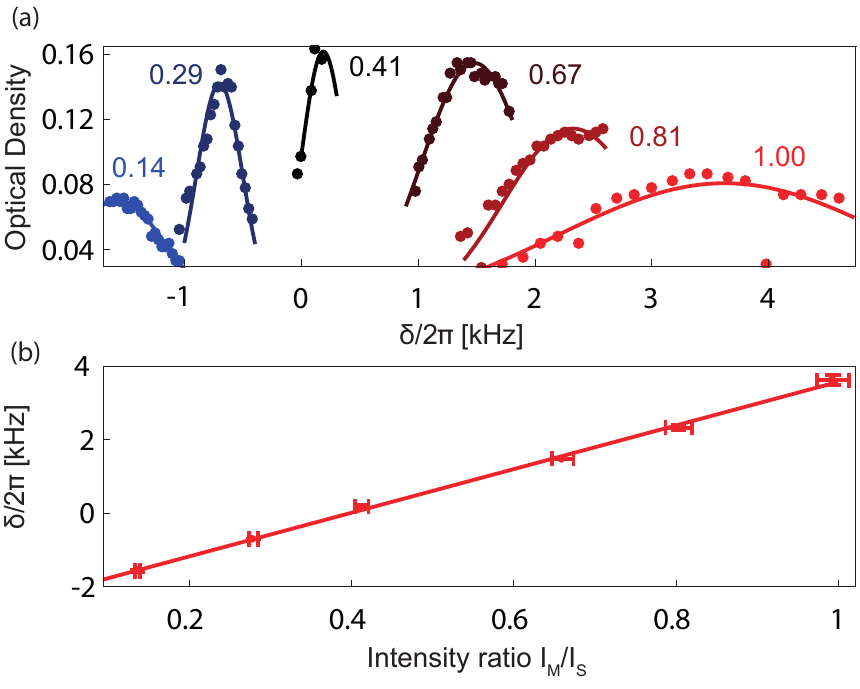}
\caption{\label{fig:shifts}(a): Raman spectra taken for different intensity ratios $R=I_M/I_S$, which are shown next to the corresponding spectrum. A clear shift in frequency is visible, as well as a broadening effect for ratios different than $R=0.40$. The spectra are fitted to extract the center frequency. (b): The center frequencies are plotted against the intensity ratio. The shift in frequencies is clearly a linear function of the intensity ratio as expected from Eq.~\eqref{eq:shifts}, with fit parameters $\Delta_3=-2\pi\times2781\,$MHz and $P_M=0.77\,\mu$W.}
\end{figure}

In order to examine this influence, we tune the relative frequency of the Raman lasers close to the theoretically expected value.
We reduce the laser powers to values around $1\,\mu$W, in order to minimize the effects of spectral power broadening.
At the ratio of $R=I_M/I_S=0.40$, we find spectral lines which are only a few hundreds of Hz in width (in fact the line width is pulse time limited).
Then, we intentionally change $R$ and observe the influence on the spectrum. The result is shown in Fig.~\ref{fig:shifts}. 
In this plot, the fitted center frequency for $R=I_M/I_S=0.40$ corresponds to an absolute relative Raman frequency of $\Omega_{01}=2\pi\times 6,834,678,300\pm 50\,$Hz (note: in the figures we use the relative frequency $\delta$ instead of the absolute frequency $\Omega_{01}$) as measured by the heterodyne beat signal of the two lasers.
The theoretically predicted frequency at the magic-field $B=3.229\,$G amounts to $\Omega_{01}=2\pi\times 6,834,678,113\pm 20\,$Hz. 
The discrepancy of $187\,$Hz can be ascribed to differences between our experimental trap bottom and the theoretical value, as well as to a frequency drift of our $10\,$MHz frequency reference.
We clearly see a shift of the spectral lines with changing the ratio, as well as a broadening effect for ratios different than $R=0.40$. 
The broadening effect results from the Gaussian intensity distribution of the exciting laser beams.\par

If we fit the spectra to obtain the center frequencies, and plot these frequencies against the ratios (see Fig.~\ref{fig:shifts}), we obtain a linear dependency as expected.
We fit the expression, which we obtained from Eq. \eqref{eq:shifts} by setting $\Delta E_0-\Delta E_1=0$, to the data. 
We use $\Delta_3$ and the power of the master laser $P_M$ (which is trivially related to the peak intensity for a Gaussian beam) as free parameters.
From the fit, we obtain $\Delta_3=-2\pi\times (2781\pm53)\,$MHz and $P_M=0.77\pm 0.01\,\mu$W. 
The first is in good agreement with our locking point at $\Delta_3=-2\pi\times 2792\,$MHz, the latter is consistent with the laser power chosen for this measurement. 
The measurements clearly confirm the predicted influence of the light shifts.
\subsection{\label{ch:asym}Asymmetric line shape}
\begin{figure}
\includegraphics{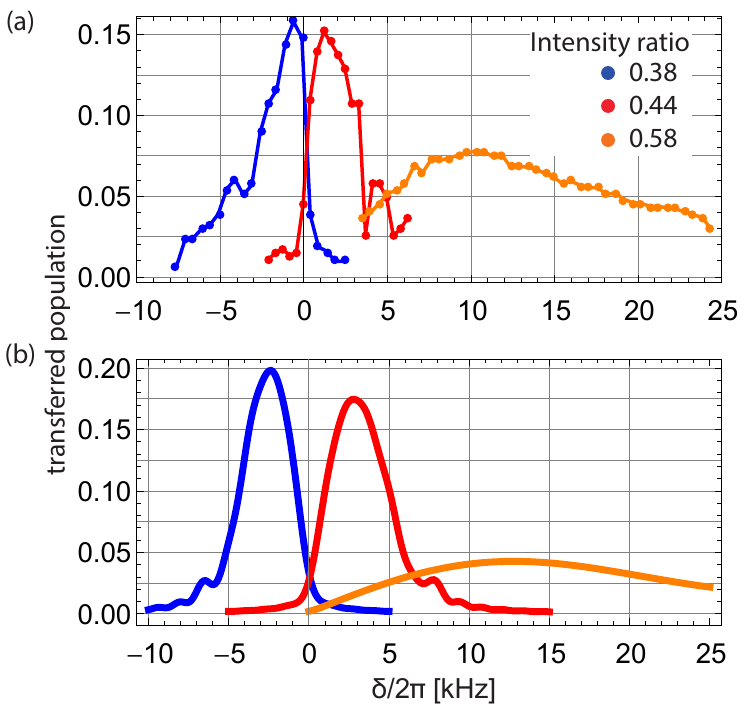}
\caption{\label{fig:asymmetry}(a): Relative atomic population transferred to $\ket{1}$ after a Raman excitation pulse of $5\,$ms taken for different intensity ratios $R=I_M/I_S$ at overall higher intensities compared to Fig.~\ref{fig:shifts}. For ratios different from $R=0.40$ we see an asymmetric broadening to either negative or positive frequencies for lower or higher ratios, respectively. (b): Simulated spectra for the same intensity ratios, $P_M=25\,\mu$W and a $5\,$ms excitation pulse. The simulation is based on the dynamics of the effective two-level system $\{\ket{0},\ket{1}\}$ for a Gaussian intensity profile of the excitation beam. Here, $\delta=0$ refers to the frequency offset at the magic field and for vanishing differential light shifts.}
\end{figure}

One distinctive feature of the Raman spectra in Fig.~\ref{fig:shifts} is a clear broadening effect for ratios different from $R=0.40$. 
When we perform the measurements with higher laser power we see two different effects: (1) a clear broadening of the spectral lines, even for $R=0.40$, and (2) an asymmetric line shape. 
The first effect can be explained by power broadening, the second needs a more elaborate explanation.
In Fig.~\ref{fig:asymmetry} we show three experimental lines for $R=0.38$, 0.44 and 0.58, showing the relative atomic population transferred to $\ket{1}$ after a Raman pulse of $5\,$ms.
The relative population is obtained by referencing the atoms detected in $|1\rangle$ to the overall atom number in the magnetic trap. 
For ratios $R<0.40$ we see a clear asymmetric lineshape with a tail towards negative frequencies, for $R>0.40$ the asymmetry is shifted to positive frequencies. 
Again, the spectrum at $R=0.58$ is significantly broadened compared to the other ratios closer to $R=0.40$.\par
In order to get a better understanding of the underlying effect, we take $H_\text{coh}$ from Eq. \eqref{eq:master} and reduce it to an effective Hamiltonian $H_\text{coh}^\text{eff}$ describing the dynamics of the two-level system constituted by the states $\ket{0}$ and $\ket{1}$.
Using this approximation we numerically solve the Schr{\"o}dinger equation $i\hbar\dot{\Psi}(t)=H_\text{coh}^\text{eff}\Psi(t)$ for different relative frequencies of the Raman lasers.
Assuming $\Psi(t)=\ket{0}$, we extract $|\langle 1\ket{\Psi(t)}|^2$ at $t=5\,$ms.
The theoretical spectra obtained this way just show a frequency shift for different ratios $R$, but no asymmetric feature.

The asymmetric lineshape can only be explained if we account for the fact that our excitation laser has a Gaussian intensity profile with a waist $w=100\,\mu$m.
We solve the same Schr\"odinger equation as before, but this time we average the results over the Gaussian intensity profile of the excitation laser and the distribution of the atoms in the magnetic trap.
Theoretical spectra obtained this way indeed feature an asymmetric lineshape for $R\neq0.40$, with an increasing asymmetry for longer pulse times.
The asymmetric tail appears consistently at negative frequencies for $R<0.40$, and at positive frequencies for $R>0.40$, which can be qualitatively understood from the light shifts. 
As a comparison to the experimental data we plot the theoretical spectra for $R=\{0.38,0.44,0.58\}$, a master laser power of $P_M=25\,\mu$W and $t=5\,$ms in Fig.~\ref{fig:asymmetry}.
Even though the theoretical description is simplified, the data is qualitatively well described. We reproduce the asymmetry, the overall frequency shift and the difference in transferred atomic population.
This measurement again confirms the significance of the optimal intensity ratio $R=0.40$.
Furthermore, the asymmetry can clearly be ascribed to the non-uniform intensity distribution of the excitation laser.
\subsection{\label{ch:temporal}Temporal dynamics}
Ideally, in the context of quantum information, one aims at the fully coherent transfer of atomic population between $\ket{0}$ and $\ket{1}$.
As discussed in section \ref{ch:full}, the dynamics of our system are not fully coherent due to off-resonant scattering [Eq. \eqref{eq:lindblad}] and trap loss [Eq. \eqref{eq:loss}].
Furthermore, the non-uniform intensity of our Gaussian laser beams will lead to additional dephasing.\par
\subsubsection{Trap loss}
\begin{figure}
\includegraphics{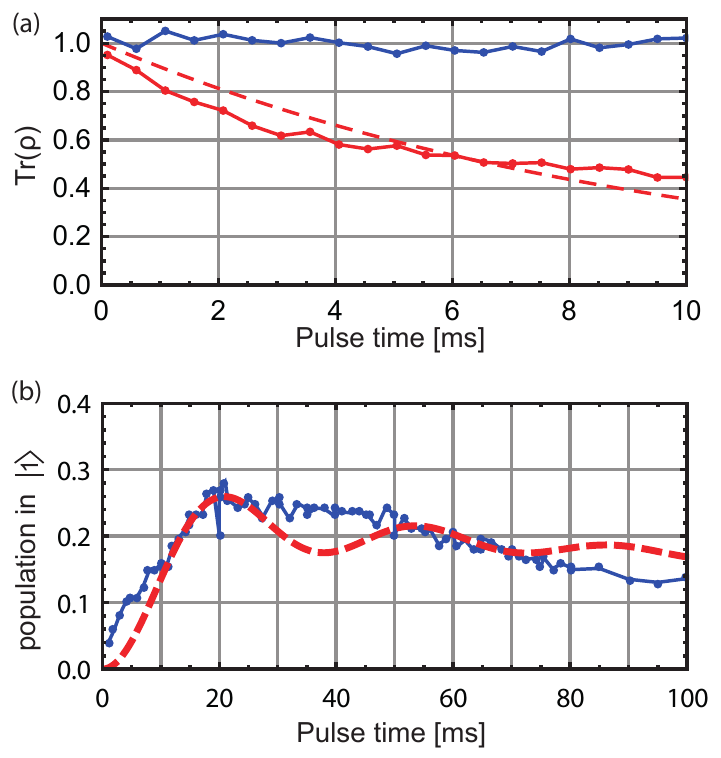}
\caption{\label{fig:temporal}(a): Total atomic population detected in the magnetic trap after a Raman pulse with varying pulse length. The blue data set is measured without the Raman lasers and shows the typical fluctuations of detected atoms in the magnetic trap. The red data set is measured using $100\,\mu$W for the master and slave laser power, and shows a decay in the amount of detected atoms with increasing pulse time. This loss results from atoms being optically pumped into non-trappable states. The red, dashed line shows a theoretical fit of Tr($\rho$) for $\gamma_\text{loss}=2\pi\times164\,$Hz.
(b): Atomic population in the $\ket{1}$ state for varying pulse length of the Raman lasers, at the optimal power ratio. The red, dashed line is a simulation assuming a master laser power of $4\,\mu$W.}
\end{figure}

In order to get a better quantitative understanding, we need to specify the parameter $\gamma_\text{loss}$ in Eq. \eqref{eq:loss}. 
This parameter empirically describes the rate at which atoms in non-trappable states leave the magnetic trap.
During the Raman excitation, these states are populated by off-resonant scattering through excited levels.
We can obtain an estimate for $\gamma_\text{loss}$ by measuring the total number of atoms in the trap after the Raman excitation pulse. 
Therefore, using the repump laser, we transfer all the atomic population to the $F=2$ ground state before taking an absorption image.
As a reference, we perform this measurement while blocking the Raman lasers, giving a nearly constant number of atoms (blue data in Fig.~\ref{fig:temporal}) with time.
The number of atoms detected is referenced to the mean value of this data.
The line features typical fluctuations due to a changing number of atoms in the magnetic trap and noise in the imaging sequence.
Then we perform the measurement using each of the Raman lasers at a power of $100\,\mu$W.
Clearly, atomic population is lost from the trap with increasing pulse length (red data in Fig.~\ref{fig:temporal}).\par

As the total atomic population is equivalent to the trace Tr($\rho(t)$) of the system's density matrix, we can directly compare our measurement to the theoretical predictions of Eq. \eqref{eq:master}.
Then, numerically solving Eq. \eqref{eq:master} for $\rho(0)=\ket{0}\bra{0}$ and varying parameters $\gamma_\text{loss}$, we fit Tr($\rho(t)$) to the data. 
The resultant best fit is shown in Fig.~\ref{fig:temporal}(a) for $\gamma_\text{loss}=2\pi\times164\,$Hz.
This loss rate is in the same order of magnitude as the larger of the trapping frequencies of our magnetic trap.
The fit describes the data well, despite the fact that we neglected the non-uniform intensity of the Raman lasers for computational reasons.

\subsubsection{Population transfer}
After having extracted an estimate for $\gamma_\text{loss}$, we can give a full theoretical account of the temporal evolution of $\rho(t)$ by solving Eq. \eqref{eq:master} for $\rho(0)=\ket{0}\bra{0}$ and compare it to the data. 
We measure the atomic population transferred to $\ket{1}$ as a function of the Raman pulse length at the optimal intensity ratio. 
Instead of a coherent Rabi flopping between $\ket{0}$ and $\ket{1}$, we see a build-up of population in $\ket{1}$ with a subsequent decay for longer pulse times [see Fig.~\ref{fig:temporal}(b)].
The fact that we observe the maximum population transfer at $20\,$\,ms confirms the effective reduction of the Raman Rabi-frequency as described by Eq.~\eqref{eq:rabi}.\par

In principle, we obtain a theoretical prediction by solving Eq.~\eqref{eq:master} and extracting $\bra{1}\rho(t)\ket{1}$.
However, we have to account for the non-uniform intensity of the Raman lasers. 
In order to do so, we first simplify the problem by assuming a one-dimensional atomic distribution.
This simplification is justified by the fact that our trap is very elongated along the axial trap dimension ($\sigma_x=266\,\mu$m, $\sigma_{y,z}=23\,\mu$m), and that the radial dimensions are smaller than the beam waist $w=100\,\mu$m.
We numerically solve Eq. \eqref{eq:master} for a number of discrete distances $x$ from the beam center, with a laser intensity of $I(x)=I(0)\cdot \exp(-2 x^2/(100\,\mu\text{m})^2)$.
Here, $I(0)$ is the peak intensity in the beam center. 
Then we weight the resulting $\bra{1}\rho(t)\ket{1}$ with the atomic distribution $\propto \exp(-x^2/2\,(266\,\mu\text{m})^2)$ and average the results over the discrete set of $x$. 
The resultant curve for a master laser power of $4\,\mu$W is shown in Fig.~\ref{fig:temporal} (b).
The initial build-up and subsequent decay is well described by our model, as well as the maximally transferred population.
We do not measure the slightly oscillatory behavior predicted by the theory.
This is most likely caused by fluctuations in our system, especially in the relative laser intensity as discussed in section \ref{ch:RamanDiscussion}.
Furthermore, at timescales of tens of ms, the atoms travel significant distances in the magnetic trap, probing areas of different intensities.

\section{\label{ch:RamanDiscussion}Discussion}
\begin{figure}
\includegraphics{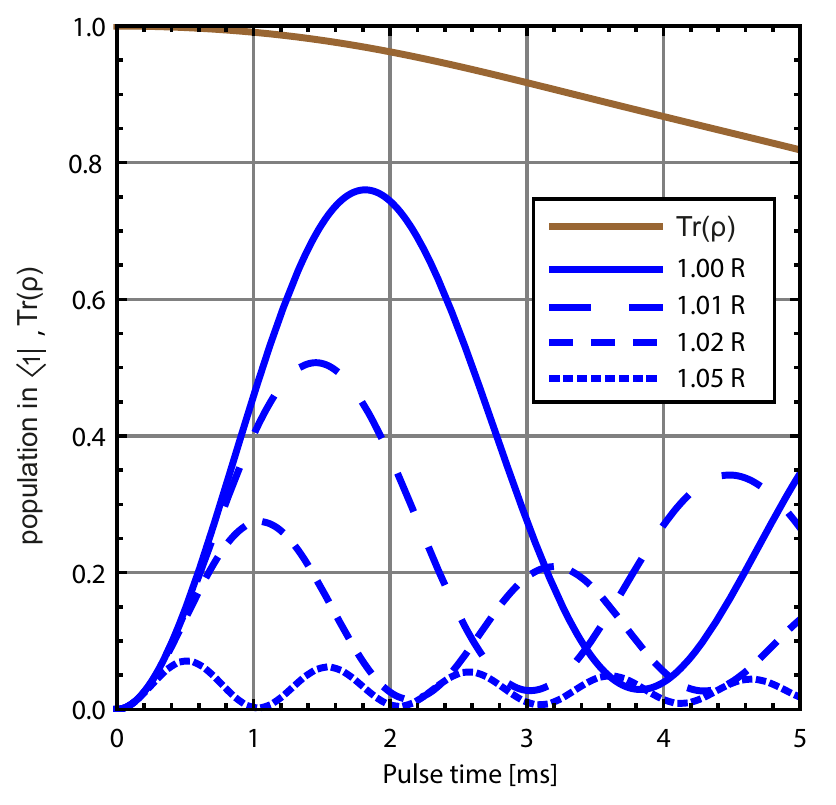}
\caption{\label{fig:uniform}Simulated temporal evolution of the atomic population in $\ket{1}$ assuming a uniform intensity distribution of the Raman laser beams at $P_M=35\,\mu$W. The simulation is performed for different intensity ratios, expressed as $\{1.00,1.01,1.02,1.05\}$ times the optimal ratio $R$. It reveals that even a small deviation from the optimal ratio leads to a significant reduction of the Rabi flopping between $\ket{0}$ and $\ket{1}$. As a comparison, the total remaining atomic population is shown.}
\end{figure}

Many of our observations, such as the asymmetric lineshape and the temporal behaviour, can be ascribed to the non-uniform intensity distristribution of the lasers.
Furthermore, the net effect is unwanted decoherence with respect to the Raman transition. 
In optical dipole traps \cite{Urban:2009jdbaca}, optical lattices \cite{Bloch:2005gn}, as well as in our magnetic microtraps \cite{Leung:2014gw}, atoms can be strongly confined such that they experience a near-uniform intensity distribution even for a Gaussian beam.
For such a uniform intensity distribution, we can simulate the Raman transition by solving Eq.~\eqref{eq:master}.
The temporal evolution of the atomic population in $\ket{1}$ is plotted for a master laser power of $P_M=35\,\mu$W and the optimal intensity ratio in Fig.~\ref{fig:uniform}. 
The population undergoes coherent Rabi flopping between $\ket{0}$ and $\ket{1}$, but never exceeds 80\% population in $\ket{1}$ due to off-resonant scattering and trap loss.
Hence, achieving a transfer of atomic population with an efficiency needed for quantum information experiments is not possible in this arrangement. 
Furthermore, the coherent transfer is highly sensitive to relative fluctuations in the laser intensities due to the induced light shifts.
As visible in Fig.~\ref{fig:uniform}, relative changes of a few percent significantly deteriorate the population transfer.
For a relative change in power of 5\% almost no atomic population is transferred to $\ket{1}$.
A possible solution is to stabilize the intensity of the Raman lasers or take both frequencies from the same laser source as in reference \cite{Yavuz:2006gj}.
\par

The issue of off-resonant scattering can be mitigated by changing the excitation scheme to the D1-line.
As evident from Eq.~\eqref{eq:rabi}, the larger splitting $\Delta_{21}$ between the $F=1$ and $F=2$ excited states ($817\,$MHz compared to $157\,$MHz for the D2 line) will lead to a relative increase of the Rabi frequency compared to the scattering rate. 
For example, for a detuning of $\Delta_1=-2\pi\times 3000\,$MHz to the $F=1$ level, the difference amounts to a factor of $\sim 4$.
Alternatively, one could resort to a scheme involving two-photon excitation with MW and RF radiation in combination with optical addressing as in reference \cite{Weitenberg:2011gn}.

\section{\label{ch:RamanConclusion}Conclusion}
Two important aspects of the Raman transition are the effective reduction of the Rabi-frequency due to the two-path interference (see Eq. \eqref{eq:rabi}), and the light shift caused by the off-resonant coupling to the excited state manifold (see Eq. \eqref{eq:shifts}). 
The differential light shift between $\ket{0}$ and $\ket{1}$ is linear in the relative intensity of the Raman lasers, which is confirmed by our measurements (Fig.~\ref{fig:shifts}).
Furthermore, there is an optimal ratio $R=I_M/I_S$ of the laser intensities for which this relative light shift vanishes. 
For our detuning, $\Delta_3=-2\pi\times2792\,$MHz, we find $R=0.40$ which is again consistent with our measurements (Fig.~\ref{fig:shifts} and Fig.~\ref{fig:asymmetry}).
The light shift in conjunction with the non-uniform intensity of the lasers leads to an asymmetric lineshape, if we deviate from the optimal ratio (Fig.~\ref{fig:asymmetry}).
This asymmetry is well described by a two-level approximation to Eq.~\eqref{eq:master} and the integral over the Gaussian beam intensity.
Instead of a coherent Rabi flopping between $\ket{0}$ and $\ket{1}$, we see incoherent build-up of atomic population in $\ket{1}$ with a subsequent decay.
Maximally 25\% of the atomic population is transferred to $\ket{1}$.
This can be explained by solving Eq.~\eqref{eq:master} for different intensities in the Gaussian beam
and performing an average.\par
Even for the case of uniform laser intensity we see from Fig.~\ref{fig:uniform} that we cannot achieve full transfer of population.
We are limited by the influence of off-resonant scattering, which can, in principle, be reduced by changing the excitation scheme to the D1 line.
From Fig.~\ref{fig:uniform} we can also conclude that the Raman transition is highly sensitive to relative intensity fluctuations of the lasers.
The combination of stable laser intensities and the excitation scheme via the D1 line could yield a significant improvement for the Raman transfer of the magnetic clock states.
\par

\begin{acknowledgments}
We thank Daniel Nicolai for his contributions to the experimental setup.
Our work is financially supported by the Foundation for Fundamental Research on Matter (FOM), which is part of the Netherlands Organisation for Scientific Research (NWO).  
We also acknowledge financial support by the EU H2020 FET Proactive project RySQ (640378).
JN acknowledges financial support by the  Marie Curie program ITN-Coherence (265031). 
\end{acknowledgments}


\bibliography{raman}

\end{document}